\documentclass[showpacs,superscriptaddress,twocolumn]{revtex4-1}%
\usepackage{hyperref}
\usepackage{graphicx}
\usepackage{amsfonts}
\usepackage{amssymb}
\usepackage{amsmath}

\begin{document}%

\title{Preisach images of a simple mechanical system}

\author{M.W.~\surname{Gutowski}}
\email [Corresponding author: ]{marek.gutowski@ifpan.edu.pl}
\affiliation{Institute of Physics, Polish Academy of Sciences,
 Warszawa, Poland}
\author{Amy \surname{Chyao}}
\affiliation{student, Plano East Senior High School, Plano, Texas}
\author{S.~\surname{Markovskyi}}
\affiliation{Institute of Physics, Polish Academy of Sciences,
 Warszawa, Poland}

\begin{abstract}
This work is an an~early stage of a larger project aiming at answering
the question whether or not the Preisach map is really fingerprinting
magnetic materials.\  More precisely, we are interested whether Preisach
model of magnetic hysteresis indeed contains any physics or is just
a~convenient modeling tool.\ To this extent we study a~very simple
mechanical system, thus fully tractable, subjected to the
external force.\  Despite of its simplicity, our model captures
all the fundamental features of real magnetic materials, namely their
hysteretic behavior, coercivity, remanent magnetization and saturation
at high fields.\
Both the overall shape of major hysteresis loop as well as First Order
Reversal Curves (FORC's) are reproduced quite correctly; they are very
similar to those observed in magnetic materials.\
The model essentially consists of a~single, spring loaded, rigid and
rotative bar with non-zero friction torque.\  The length of a~projection
of this bar onto the direction of an external force is identified with
magnetization.\  The friction torque and the spring constant are the
only freely adjustable parameters of our model.\  Here we investigate,
and present, their influence on the inferred Preisach maps.
\end{abstract}

\pacs{
07.05.Tp, 
45.20.da, 
75.60.Ej  
}
\maketitle

\section{Motivation}
It took more than ten years to put into experimental practice
\cite{Pike} the idea of the so called FORC's (First Order Reversal
Curves) first introduced by Mayergoyz \cite{IM} in 1986.\ Since then
the FORC diagrams became more or less standard method of analysis of
magnetic materials exhibiting hysteretic behavior.  The FORC diagrams
are nearly identical with Preisach \cite{Preis} maps, except for the
boundary part (necessarily equal to exactly zero in FORC picture),
corresponding to the so called reversible part of magnetization curve.\
For this reason we will be using both terms as if they were equivalent.

In the beginning, the shape of Preisach map was essentially guessed.
According to some theoretical hints given by N\'eel (1949--1952), such
a~map should consist of a single peak.\ The parameters of this peak,
assumed most often as being Gaussian, Lorentzian or lognormal, were
later refined in such a way as to reproduce the experimental data. The
FORC technique, in contrast, does not assume any particular shape of
a~Preisach map and therefore reproduces empirical facts more
accurately.\ It is really good news for engineers, but physicists rather
would be interested what is the meaning of various details observed in
result of FORC-type analysis.\ More specifically, it is very
important to be able to draw some definite conclusions concerning
various magnetic interactions dominating in a~given material.

The published data with recovered FORC diagrams are quite ubiquitous
but lack systematic attitude.  The experimental data are sometimes
accidental (rock samples), or only a very limited set of samples is
available.\ In addition, the experiment itself is rather time-consuming.
On the other hand, simulated data are easy to generate and, what is
even more important, the analyst `knows everything' about her artificial
`sample' in advance.\ Manipulating various parameters is straightforward
and practically unrestricted.

In this situation, it seems natural to develop systematic studies
in that matter, starting from the simplest models and making them
more realistic (and more complex) later.\ Here we present our first
attempt in this direction.

\section{The model}
\begin{figure}[h]
    \includegraphics[keepaspectratio,angle=0,width=0.5\hsize]{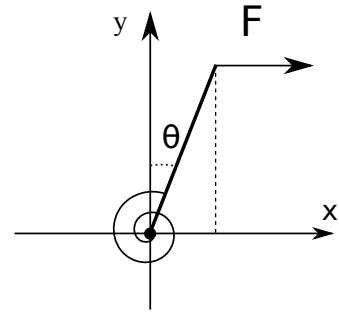}
    \caption{A~sketch of the mechanism imitating magnetic system.\ In the
    absence of friction, the equilibrium position of a~rotary rod
    (thick line) is at $\theta=0$.\ The external force $\vec{F}$ is
    always parallel to the $x$-axis.
    }\label{arrow}
\end{figure}

In our model (see Fig.~\ref{arrow}) a~stiff rod can be rotated around the axis
located at (0,0).\ The rod is connected with the rotation axis by means of
a~spring.\
An external force is applied either parallel or antiparallel to the $x$-axis,
so the rod's deflection $\theta$ from vertical direction may never exceed $\pi/2$.\
A~static friction between the rod and its rotation axis is present.\
No movement is possible whenever the torque produced by spring and the external
force has lower magnitude than the static friction torque, i.e. when
\begin{equation}
\left|\tau_{\textrm s} + \tau_{\textrm F}\right|
\leqslant \left|\tau_{\textrm f}\right|,\label{move}
\end{equation}
where the spring torque $\tau_{\textrm s} = -k\theta$,
with the dimensionless parameter $k$ being called \emph{spring constant}
from now on, and the external force produced torque is\
$\tau_{\textrm F}=(\vec{\mu}\times{\vec{F}})_{z}=
\mu\,\sin(\theta-\frac{\pi}{2})=\mu\,\cos\theta$, where $\mu\,(\equiv{1})$
is the length of a~rod, and, finally, $\tau_{\mathrm{f}}$ is the static
friction torque.

When the relation (\ref{move}) is not satisfied, the rod rotates in the
direction dictated by the sign of $\tau_{\textrm s}+\tau_{\textrm F}$
until the equilibrium condition
\begin{equation}
\left|\tau_{\textrm s}+\tau_{\textrm F}\right|
-\left|\tau_{\mathrm{f}}\right| = 0
\end{equation}
is reached.

In our mechanical model of magnetic hysteresis the external force $F$
is identified with the exciting field $H$, and the projection of a~rod
onto the direction of external field, equal to $\mu\,\sin\theta$,\
corresponds to (reduced) magnetization, since $\mu\equiv{1}$.

\section{The results}
\begin{figure}[h]
    \includegraphics[keepaspectratio,angle=-90,width=0.8\hsize]{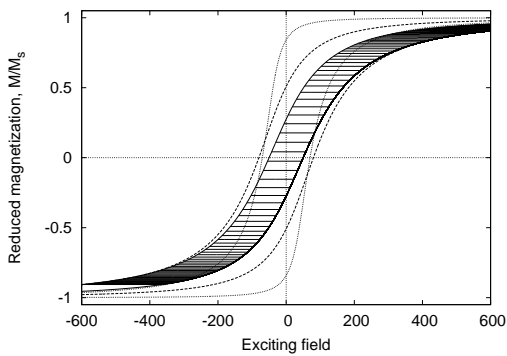}
    \caption{Hysteresis in the system under study. Parameters (spring constant,
    friction) are: (70,70) for dotted loop, (150,80) for dashed one, and
    (180,50) for the loop with 99 first order reversal curves shown.
    }\label{loops}
\end{figure}

\begin{figure}[h]
    \includegraphics[keepaspectratio,angle=0,width=\hsize]{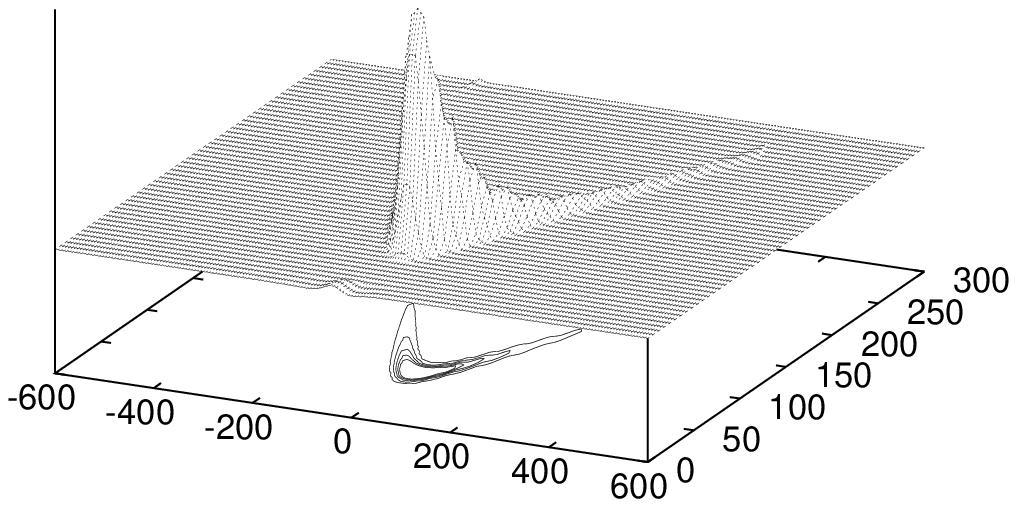}
\vskip -2cm
    \includegraphics[keepaspectratio,angle=0,width=\hsize]{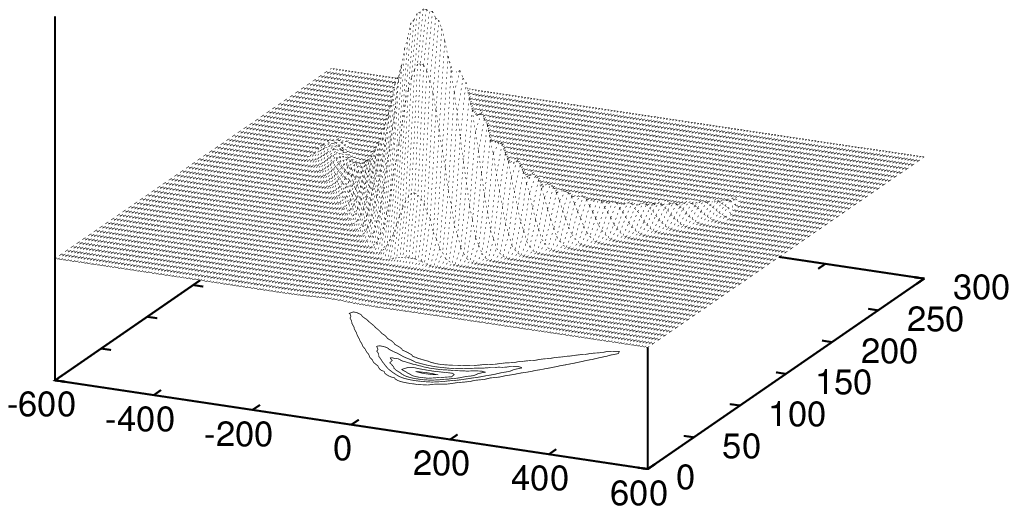}
\vskip -2cm
    \includegraphics[keepaspectratio,angle=0,width=\hsize]{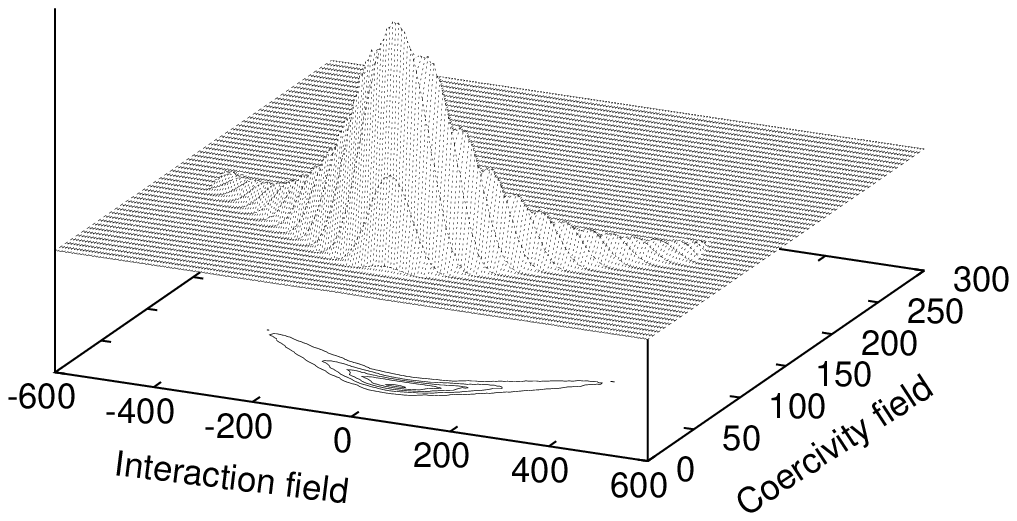}
    \caption{FORC diagrams for the same combinations of spring constant
    and friction as in Fig.~\ref{loops}.  From top: $(70,70)$, $(150,80)$,
    and $(180,50)$.\ Note: the $z$-coordinate is scaled to keep the height
    of peaks fixed, but noisy contour $z=0$ is omitted to show more
    clearly the peak's shape.
    }\label{ex3}
\end{figure}

We present three selected hysteresis curves in Fig.~\ref{loops}.\
The shape of FORCs,
perfectly horizontal, may be a~little surprising at first sight, since
our FORCs differ significantly from the majority of published data.
Yet, contrary to intuition, our simulations are not completely
unrealistic.\ One might expect horizontal reversal curves only for very
hard magnetic materials, but the curves similar to ours have been
recently observed for arrays of Ni microwires \cite{Ni-wires}
electrodeposited in nanoporous alumina templates.\
The overall shape of our recovered FORC diagrams is roughly the same for
sensible combinations of adjustable parameters.\ Three examples are shown in
Fig.~\ref{ex3}.

It is worth noticing that the peak is not a~convex body -- note the presence
of `arms' and `folded ellipses' as contour lines.\ This fact alone
invalidates some quasi one-dimensional, simplified models of the Preisach
maps.\  It is also in sharp contrast with other analytical model of magnetic
hysteresis, proposed by Tak\'acs \cite{Takacs} and shown in Fig.~\ref{meas}.
The results of more quantitative analysis are shown in following figures.\

\begin{figure}[h]
\vskip -1cm
    \includegraphics[keepaspectratio,angle=00,width=0.99\hsize]{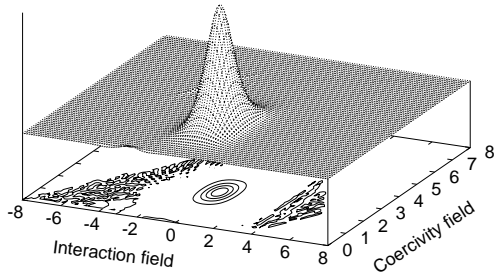}
    \caption{Preisach image for Takacs model of magnetic hysteresis
    \cite{Takacs}.\ Note the very regular peak shape.
    }\label{meas}
\end{figure}

\begin{figure}[h]
    \includegraphics[keepaspectratio,angle=00,width=0.99\hsize]{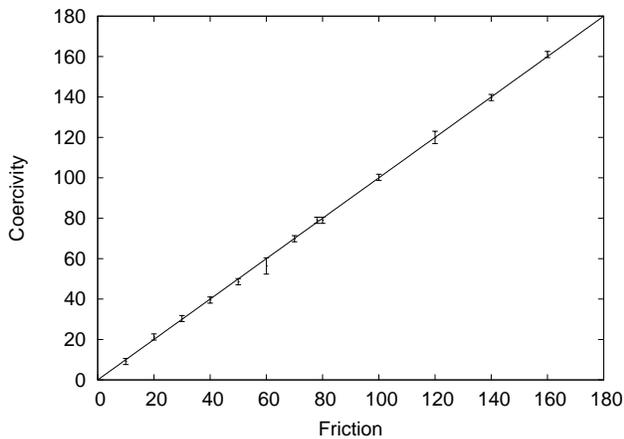}
    \caption{The coercivity is strictly proportional to friction.
    }\label{coerc}
\end{figure}

The position of the highest point on a~map, more precisely
its \emph{coercivity field} coordinate, is strictly proportional to the
friction but is insensitive to the spring stiffness (Fig.~\ref{coerc}).\
The error bars there are taken as the field values nearest to the crossing
of major hysteresis loop and $M=0$ axis.

\begin{figure}[h]
    \includegraphics[keepaspectratio,angle=00,width=0.99\hsize]{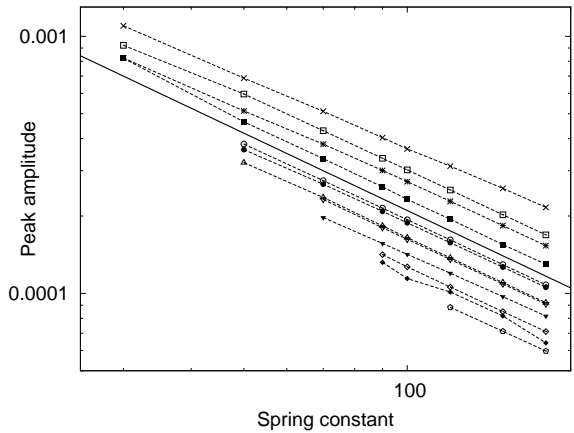}
    \caption{Main peak's amplitude \emph{vs.\/} spring constant for different
    friction constants.\ Solid line shows
    $1/(\mathrm{spring}\ \mathrm{constant})$ dependence.\ Friction is
    (from top): $10$, $20$, $30$, $40$, $50$, $60$, $70$, $80$, $100$,
    $120$, $140$, and $160$.\ The lines connecting points are eye-guides only.
    }\label{amplit}
\end{figure}

\pagebreak[4]
The peak's amplitude, on the other hand, is a~decreasing function of both
parameters.\ It seems inversely proportional to the spring constant
(see Fig.~\ref{amplit}).
The dependence on friction is less clear, due to inaccuracies generated
by imperfect smoothing procedure, resulting in a~slightly warped peak
shape, as one can see in Fig.~\ref{ex3}.\ Nevertheless, it is quite likely
that the peak's height is inversely proportional to the friction as well.

\section{Conclusions}
Our results show that various approximations of the peak shape, existing
in literature, are probably too simplistic or even incorrect.

\smallskip
We occasionally
observe small negative regions on obtained Preisach maps, especially at
their edges, near coercivity coordinate equal to zero (see the top of
Fig.~\ref{ex3}).\
We are strongly convinced that it is an artifact, due to the numerical
differentiation and/or smoothing procedure, rather than the real thing.\

Negative regions are often observed both in real and simulated \cite{Robb}
magnetic systems.  This feature is most likely
related to the multimodal character of a~potential energy landscape
in real systems (pinning sites, defects, domain structure), while in our
model the potential energy is unimodal.

\section*{Acknowledgments}
This work was partially supported by the European Union within
the European Regional Development Fund, through the Innovative
Economy grant (POIG$.01.01.02-00-108/09$).


\end{document}